\documentclass[12pt,letterpaper]{article}
\usepackage{epsfig}
\usepackage{setspace} 
\usepackage{footmisc}
\usepackage{graphicx}
\usepackage{amsfonts}
\usepackage{amsmath}
\usepackage{amsthm}
\usepackage{float}
\usepackage{color}
\usepackage{natbib}
\usepackage{verbatim}
\usepackage
[
	papersize={8.5in,11in},
	hmargin=1.25in,
        top=1.25in,
        bottom=1.25in
]
{geometry}

\RequirePackage{bm}

\newcommand{\footremember}[2]{%
   \footnote{#2}
    \newcounter{#1}
    \setcounter{#1}{\value{footnote}}%
}
\newcommand{\footrecall}[1]{%
    \footnotemark[\value{#1}]%
} 
\title{Data Mining to Investigate the Meteorological Drivers for Extreme Ground Level Ozone Events}
\author{%
    Brook T. Russell\footremember{clemson}{Clemson University, Department of Mathematical Sciences}%
    \and Daniel S. Cooley\footremember{csu}{Colorado State University Department of Statistics}%
    \and William C. Porter\footremember{mit}{Massachusetts Institute of Technology Department of Civil and Environmental Engineering}%
    \and Brian J. Reich\footremember{ncsu}{North Carolina State University Department of Statistics}%
    \and Colette L. Heald\footrecall{mit} \footnote{Massachusetts Institute of Technology Department of Earth, Atmospheric and Planetary Sciences}%
}

\begin{document}
\maketitle


{\bf Keywords:}  tail dependence, multivariate regular variation, constrained optimization, cross validation, smooth threshold

\begin{abstract}
This project aims to explore which combinations of meteorological conditions are associated with extreme ground level ozone conditions. 
Our approach focuses only on the tail by optimizing the tail dependence between the ozone response and functions of meteorological covariates.
Since there is a long list of possible meteorological covariates, the space of possible models cannot be explored completely.
Consequently, we perform data mining within the model selection context, employing an automated model search procedure.
Our study is unique among extremes applications as optimizing tail dependence has not previously been attempted, and it presents new challenges, such as requiring a smooth threshold. 
We present a simulation study which shows that the method can detect complicated conditions leading to extreme responses and resists overfitting. 
We apply the method to ozone data for Atlanta and Charlotte and find similar meteorological drivers for these two Southeastern US cities. 
We identify several covariates which help to differentiate the meteorological conditions which lead to extreme ozone levels from those which lead to merely high levels.
\end{abstract}

\section{Introduction and Motivation}
Ground level ozone (O$_3$) is known to be detrimental to the human respiratory system \citep[Section~5.2]{EPA2006}.  
Research indicates that acute exposure to ozone can lead to a decline in lung function and increased inflammation \citep[Section~7.2.8]{EPA2006}.  \cite{bell2004} report that there is a relationship between increases in ozone and mortality in urban areas.  
\cite{wilson2014} find that the effect of ozone is non-linear and that extremely high levels of ground level ozone could be especially harmful.  
For these reasons, it is important to understand what are the contributing factors which lead to the most extreme ozone levels.

\begin{figure}[h]
\begin{center}
\includegraphics[width=0.9\textwidth]{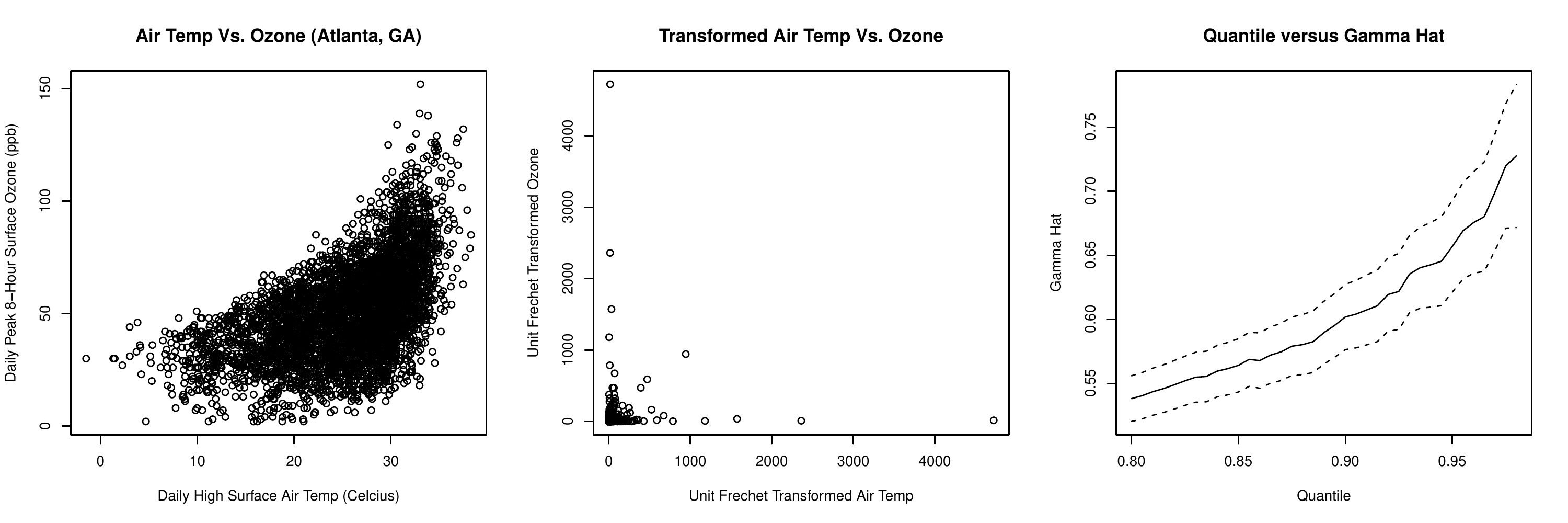}
\caption{(L) A scatterplot of daily high surface air temperature versus peak daily maximum eight-hour surface ozone in Atlanta, Georgia from 1992 to 2010 (April-October).  (C) A scatterplot of the same variables after transforming to the unit Fr\'echet scale via rank transformations.  (R) A plot of $\hat{\gamma}$, as defined in Equation (\ref{eq:defGammaHat}), for a sequence of quantiles for air temperature and ozone with 95\% confidence bands.}
\label{fig:tempVSozone}
\end{center}
\end{figure}

Ozone is a secondary pollutant, 
created via a chemical reaction which occurs when nitrogen oxides ($\text{NO}_{\text{x}}$) and volatile organic compounds (VOCs) are exposed to ultraviolet radiation from sunlight.  
The meteorological drivers which are associated with high ozone levels (high temperature, low wind speed, high solar radiation) are well known \citep{jacob2009}.  
However, it is less well known what meteorological conditions distinguish an extreme ozone day from one with merely high ozone levels. 
The left panel of Figure~\ref{fig:tempVSozone} partially illustrates this idea.  
Ground level ozone is plotted versus air temperature for Atlanta, Georgia from 1992 to 2010 (April-October).  
This scatterplot shows that extreme levels of ozone occur when the air temperature is high; however, days with the highest ozone readings do not correspond to the days with the highest temperatures.  
Motivated by a larger US EPA funded project which aims to understand how atmospheric chemistry models represent extreme ozone, this study aims to better understand the meteorological drivers of extreme ozone.

We aim to find functions of meteorological covariates that have a high degree of {\em tail dependence} with ground level ozone. 
That is, we want to find functions of covariates which tend to be very large when ground level ozone is extreme. 
As is typical for an extreme value (EV) analysis, we only analyze data which are considered to be extreme and disregard that which is non-extreme.  
%
Our approach consists of two linked tasks. 
The first is an optimization problem: for a specific set of covariates (which may be transformed from or functions of the original meteorological covariates), we want to find the coefficients in the linear combination of meteorological covariates that optimize tail dependence with ozone. 
The second is a data mining problem: we aim to find which of many possible meteorological covariates are associated with extreme ozone conditions. 
Here, data mining is a model selection problem where the model space is too large to search exhaustively.
We perform data mining in a series of steps, which concludes with an automated search of the model space.

Importantly, our goal is not prediction of ozone levels.
Ozone prediction is best done by atmospheric chemistry models which capture the known physics and chemistry in terms of the differential equations which underlie these models.
Our motivation is that current models tend to poorly predict the most extreme events.
Thus our goal is exploratory:  by focusing only on extreme events, we aim to extract a signal between the extreme ozone responses and the associated meteorological conditions.
When performing the second task of model selection, we will not limit our attention to the one model which fits best, but instead will explore the characteristics and commonalities of a number of the best fitting models.
We do this for a couple of reasons.
First, we do not believe that the true relationship between meteorology and ground level ozone lies within the model space we are able to explore.
Second, we believe it is likely that there is more than one formula of meteorological conditions which can lead to extreme ground level ozone.
We model the bivariate relationship between a linear combination of (functions of) covariates and ozone via the framework of bivariate regular variation.
Defined only in terms of the joint tail, multivariate regular variation allows one to model multivariate threshold exceedances, thus focusing only on extreme behavior.
We employ a bivariate tail dependence measure to quantify the dependence between the function of covariates and the response, and this measure is optimized subject to a constraint which imposes a marginal condition required by the regular variation framework.
Our study is quite different in aim from a typical multivariate extremes study.  
The goal of most multivariate extremes analyses is to assess risk, and the quantity of interest is the estimated probability of an 
extreme event occuring simultaneously for multiple responses.  
We are unaware of any previous work which use extremes methods to optimize dependence or perform data mining.

When modeling a response in terms of covariates, it is common to consider a type of regression analysis. 
Standard linear regression models the expected response (and thus the conditional distribution's center), and consequently it tends to be a poor method for describing extremes.  
Approaches such as quantile regression or logistic regression can be tailored to focus on large values of the response.  
Our EV approach focuses on only the most extreme values of the response and is fundamentally different from regression approaches.
However, one can make an analogy between our approach and standard least-squares regression:
standard regression aims to find the linear combination of 
covariates which optimizes correlation with the response (in terms of $R$-squared), and our approach aims to find the coefficients which optimize a measure of limiting tail dependence.


Our method also differs from conditional or regression approaches for extremes \citep[Ch. 7]{Beirlant04} which model the parameters of a univariate extremes model (e.g., generalized extreme value (GEV) or generalized Pareto distribution (GPD)) as functions of covariates.
Conditional models are typically applied when the covariate is measured on a longer time scale (e.g. annual) than the response, thus allowing the researcher to extract data (e.g. annual maxima or threshold exceedances) which are considered extreme for the particular covariate value.
Conditional models for extremes have been used in atmospheric science studies to study trends by conditioning on year, or to study the relationship with slowly-evolving climatological regimes \citep{sillmann2011, maraun2011}.
In contrast, \cite{reich2013} model daily ozone levels conditional on a daily covariate and include an extremes model for the tail, but they do so by modelling the entire distribution and limiting their investigation to a single covariate.
Our approach is specifically designed for covariates which vary on the same time scale as the response.
Importantly, there is a subtle difference between the questions answered by a conditional extremes approach and our proposed approach.
Because it models the tail conditional on the covariates, the conditional approach answers the question ``Given certain conditions, what is the extreme behavior?"
By optimizing tail dependence, our approach answers a slightly different question of ``What conditions are most strongly associated with the extreme observations?"

This paper is organized in the following manner.  
In Section~\ref{sect2} we review the concepts of bivariate regular variation and tail dependence.  
We discuss tail dependence parameters and their estimators, and discuss why some estimators are better suited for optimization.
We introduce tail dependence estimators that utilize a smooth threshold.  
In Section~\ref{sect3} we present our procedure.
Section \ref{sec4}'s simulation study demonstrates the ability of our approach to detect complicated conditions which lead to extreme behavior.  
In Section~\ref{sect4}, we apply a multistep data mining procedure to data from Atlanta, Georgia and Charlotte, North Carolina, and list  meteorological covariates which exhibit a relationship with extreme ozone levels.


\section{Extremes and Dependence in the Tail} \label{sect2}

Multivariate regular variation implies that the joint tail decays like a power function.
Since the framework is defined only in terms of the joint tail, it is useful for modeling threshold exceedances.
Importantly, the modeling framework of regular variation allows for asymptotic dependence.
If $(X,Y)$ is a bivariate random vector with common marginals, then $X$ and $Y$ are asymptotically dependent if
$
  \chi = \lim_{u \rightarrow x^+} P(Y > u \mid X > u) > 0,
$
where $x^+$ is the right endpoint of the marginal distribution's support.
Most multivariate models such as those with Gaussian dependence structure and most copula models do not allow for asymptotic dependence.
Like other models for multivariate extremes, multivariate regular variation makes assumptions about the marginal distributions in order that dependence in the tail can be described.

\subsection{Bivariate Regular Variation and Exceedances}
A random vector $\boldsymbol{Z} \in [0,\infty)^2$ is regularly varying if
there exists a sequence $b(n)$ such that $P( \| \boldsymbol{Z} \| > b(n)) \sim n^{-1}$ and 
\begin{equation}
  \label{eq: regVarDefn1}
  n P \left( \frac{\bm Z}{b(n)} \in \cdot \right) \stackrel{v}{\rightarrow} \nu(\cdot),
\end{equation}
where $v$ denotes vague convergence on $\mathbb{E} = [0, \infty]^2 \backslash \{\boldsymbol{0}\}$ and $\|\cdot\|$ is any norm \citep{Resnick2007}.  
%
A useful polar coordinate representation follows by
defining a radial component, $R=\|\boldsymbol{Z}\|$ and angular component, $\boldsymbol W=\| \boldsymbol{Z} \|^{-1}\boldsymbol{Z}$. 
$\bm Z = R \bm W$ is regularly varying if
\begin{equation} \label{eq:MVRV}
n P\left( b(n)^{-1} R > r , \boldsymbol{W} \in B \right) \stackrel{v}{\rightarrow} r^{-\alpha} H(B) \text{ as } n \rightarrow \infty,
\end{equation}
where $S = \{\boldsymbol{z} \in \mathbb{E} : \|\boldsymbol{z}\| = 1\}$ is the unit sphere under any norm,
and $H$ is a finite measure for any $H-$continuity Borel subset $B$ of $S$.
Because the right hand side of~(\ref{eq:MVRV}) is a product measure, the radial and angular components become independent in the limit.  
We can characterize the tail behavior via $\nu$, or alternatively via $\alpha$ and the angular measure $H$.
$H$ contains all dependence information, and $b(n)$ can be chosen so that $H$ is a probability measure.

\citet[Proposition 5.10]{Resnick87} shows that monotone transformations of the univariate marginal distributions do not change the fundamental nature of the tail dependence, in the sense that the domain of attraction is preserved.  
Thus, the framework can be used to model multivariate data with differing tail behavior, which may or may not be heavy-tailed.  
If the data come from $\boldsymbol{Y}$ which is not regularly varying, 
we assume that there exist probability integral transformations $T_i$ ($i=1,2$), such that $T_i(Y_i)=Z_i$, and $\boldsymbol{Z} = (Z_1, Z_2)^T$ is regular varying. 
Statistical practice for multivariate extremes typically requires transforming marginals to a common, convenient distribution under which the dependence structure is more easily described.  
When dealing with real data, 
an analyst will typically transform each margin using $\hat{T}_i$, where $\hat{T}_i$ is a probability integral transformation based on an estimated distribution function.  
We choose to transform our marginals to the unit Fr\'echet distribution which has cdf $G(z) = P(Z_i \leq z) =  \exp{\{-z^{-1}\}}$.  
The unit Fr\'echet distribution is regularly varying with $\alpha=1$, therefore our transformed data will be very heavy-tailed.  
Since $\alpha=1$, we find it convenient to use the $L_1$ norm:  $\|\boldsymbol{Z}\|_1 = Z_1 + Z_2$, as this implies $H$ is a probability measure if $b(n) \sim 2n$.

Let $\boldsymbol{Z} = (Z_1,Z_2)^T$ be a bivariate regular-varying $\alpha = 1$ random vector with common marginal distributions.
$H$ can be thought of as a probability measure on $[0,1]$.
Informally, as dependence increases, the mass of $H$ concentrates toward the center (i.e., 1/2), and consequently large realizations of $\bm Z$ will tend to occur closer to the 45-degree line.
As dependence decreases, the large observations of $\bm Z$ will tend to occur near the axes. 
If $Z_1$ and $Z_2$ are asymptotically independent \citep{ledford1996}, $H$ has mass of .5 at $\{0\}$ and $\{1\}$.
When exploring data, a scatterplot after transformation to a heavy tailed distribution can help one to understand tail dependence.
The plot in the center panel of Figure \ref{fig:tempVSozone} gives a scatterplot of air temperature versus ozone in Atlanta, where both variables are transformed to have unit Fr\'echet marginals via rank transformations.  
Due to the heavy tail,  the majority of the points have congregated near the origin, and one is left to view the behavior of the large points.
As many of the large points occur near the axes, it indicates that the level of tail dependence is relatively weak; that is, temperature alone does not describe extreme ozone conditions.

\subsection{Tail Dependence Summary Parameters and Estimators}

Within the regular variation framework, tail dependence is completely described by $\nu$ or $H$.
However, neither of these quantities is easily summarized.
In order to perform a numerical optimization, we must summarize tail dependence with a single number.  

Several summary measures given in terms of $\nu$ have been proposed.
For example, if $\bm Z$ is bivariate regularly varying, 
$$
\chi =  \lim_{u \rightarrow \infty} \frac{\nu([u,\infty] \times [u,\infty])}{\nu([u,\infty] \times [0,\infty])}.
$$
An estimator of $\chi$ can be obtained by selecting a high $u$ and replacing $\nu$ with the observed counts:
\begin{equation}
\hat{\chi}(u)= \frac{ \sum_{t=1}^{n} \mathbb{I} \{Z_{t,1}>u , Z_{t,2}>u\}  }{ \sum_{t=1}^{n} \mathbb{I}{\{Z_{t,1}>u\}} }.
\label{eqn: chiHat}
\end{equation}
\cite{Coles99b} give a similar estimator for $\chi$ which is symmetric in the sense that it also considers $Z_1$ conditioned on $Z_2$ exceeding $u$.
Other tail dependence summary measures based on $\nu$ with similar estimators have been proposed;  the extremogram of \citep{davis2009} can be viewed as a generalization of $\chi$ applied in the time series context.
Tail dependence summary measures with `counting estimators' similar to (\ref{eqn: chiHat}) cannot serve as the objective function in numerical optimization.  
In (\ref{eqn: chiHat}) for a fixed $u$, $\hat{\chi}(u)$ gives the number of points that exceed $u$ in both components divided by the number of points that exceed in the first component.  
In our optimization method, any function of meteorological covariates which results in the same number of exceedances would yield the same value of $\hat{\chi}(u)$ regardless of the exceedances' values, causing the optimization to fail.

Alternatively, summary measures can be derived from $H$.
\cite{larsson2012} consider dependence summary parameters of the form
\begin{equation}
  \rho_\kappa = \int_{[0,1]}  \kappa(w) dH(w)
  \label{eq:EDMGeneral}
\end{equation}
for bounded and continuous $\kappa: [0,1] \rightarrow [0,\infty)$.  
\cite{larsson2012} propose estimators and show consistency by relying on the intermediate asymptotics common to EVT.
Let $k := k(n)$ be a sequence such that $k \rightarrow \infty$ and $k/n \rightarrow 0$, and let 
$$
  \hat \nu_n (\cdot) := \frac{1}{k} \sum_{t = 1}^n \mathbb{I} \{ \boldsymbol Z_t/b(n/k) \in \cdot \},
$$
where 
$b$ is defined as in (\ref{eq:MVRV}).
It can be shown that 
$\hat \nu_n \stackrel{v}{\rightarrow} \nu$ \citep[Theorem~4.1]{Resnick2007} on the space of positive Radon measures on $[0,\infty]^2 \setminus \{ \boldsymbol{0} \}$.
Then, for sets in $[0,1]$, define 
$$
  \hat H^{(\text{hard})}_n(\cdot) := \frac
  	{\hat \nu_n \{ \boldsymbol z \mid \| \boldsymbol z \| > 1, z_1 \| \boldsymbol z \|^{-1} \in \cdot \}}
  	{\hat \nu_n \{ \boldsymbol z \mid \| \boldsymbol z \| > 1 \}} = 
	\frac
	{\sum_{t = 1}^n \delta^{(\text{hard})} \left( \frac{\| \boldsymbol Z_t \|}{b(n/k)} \right) 
	\mathbb{I}{ \left\{ \frac{Z_{t,1}}{\| \boldsymbol Z_t \|} \in \cdot \right\}} }
	{\sum_{t = 1}^n \delta^{(\text{hard})} \left( \frac{\| \boldsymbol Z_t \|}{b(n/k)} \right) },
$$
where $\delta^{(\text{hard})}(z) = \mathbb{I}\{ z \geq 1\}$.
The superscript denotes a `hard' threshold at 1 which is standard for extremes. 
Let $W_t = Z_{t,1}/ \| \boldsymbol Z_t \|$.
We obtain the estimator for the tail dependence measure
\begin{equation} \label{eq:EDMhat}
  \hat{\rho}_{\kappa;n}^{(\text{hard})} = \int_{[0,1]}  \kappa(w) \hat{H}^{(\text{hard})}_n(dw) 
  = \frac
      {\sum_{t=1}^n \delta^{(\text{hard})} \left( \frac{\| \boldsymbol Z_t \|}{b(n/k)} \right) \kappa(W)}
      {\sum_{t=1}^n \delta^{(\text{hard})} \left( \frac{\| \boldsymbol Z_t \|}{b(n/k)} \right)}.
\end{equation}
Because $\hat \nu_n \stackrel{v}{\rightarrow} \nu$, it follows that $\hat H^{(\text{hard})}_n \Rightarrow H$ and $\hat \rho_{\kappa;n}^{(\text{\text{hard}})} \stackrel{p}{\rightarrow} \rho$ \citep{Resnick2004}. 
Resnick (\citeyear[p.~301]{Resnick2007}) further states that if $b(n/k)$ is replaced with an estimator $\hat b(n/k)$, and $\hat b(n/k) / b(n/k) \stackrel{p}{\rightarrow} 1$, the above convergences hold.


We use a dependence summary parameter based on $|Z_1 - Z_2|$, the $L_1$ distance to the 45-degree line.
The madogram \citep{cooley06} similarly used the $L_1$ difference of variates, but assumed $\bm Z$ was max-stable rather than regular-varying.
After converting to psuedo-polar coordinates, $|Z_1 - Z_2| = |RW - R(1-W)| = R|2W-1|$.  
We require a function that is not dependent upon $R$, so we choose $\kappa(w) = |2w-1|$, noting
$$
\frac{|Z_1 - Z_2|}{Z_1 + Z_2}=\frac{R|2W-1|}{R}=|2W-1|.
$$
Define the parameter 
$$\gamma = \int_{[0,1]} |2w-1| dH(w).$$
Note that $0 \leq \gamma \leq 1$, and that a smaller value of $\gamma$ implies a higher degree of tail dependence.  
When $\gamma=1$ we have asymptotic independence, whereas $\gamma=0$ shows perfect tail dependence.
As in \cite{Resnick2004}, we can define an estimator of $\gamma$ using the estimator of $H$,
\begin{equation} \label{eq:defGammaHat}
\hat{\gamma}_n = \int_{[0,1]} | 2w - 1 | \hat{H}^{(\text{hard})}_n(dw).
\end{equation}
The right panel of Figure~\ref{fig:tempVSozone} shows estimates of $\hat \gamma_n$ between the ozone response and temperature for increasing thresholds.
As $\hat \gamma_n$ achieves a level of about 0.72 shows that these two variables appear to exhibit asymptotic dependence, but the level of dependence is relatively weak. 

\subsection{Tail Dependence Estimation with a Smooth Threshold}
As we wish to perform optimization, the hard threshold typically used in EVT is problematic as points would move back and forth across the threshold during optimization, likely not allowing the optimizer to converge.
\cite{chaudhuri2011} propose using `smooth interpretations' of discontinuous functions in numeric optimization.  
We use these techniques for EVT by replacing the hard threshold, $\delta^{(\text{hard})}$, with a smoothed one, $\delta_n^{(\text{smooth})}$, which gradually increases the weights from 0 to 1 as the radial component increases.

Let 
\begin{equation}
  \label{eq:  smoothH}
  \hat H_n^{(\text{smooth})}(\cdot) :=  \frac
      {\sum_{t=1}^n \delta_n^{(\text{smooth})} \left( \frac{\| \boldsymbol Z_t \|}{b(n/k)} \right) 
      \mathbb{I}{ \left\{ \frac{\boldsymbol Z_t}{\| \boldsymbol Z_t \|} \in \cdot \right\}}}
      {\sum_{t=1}^n \delta_n^{(\text{smooth})} \left( \frac{\| \boldsymbol Z_t \|}{b(n/k)} \right)},
\end{equation}
where $\delta_n^{(\text{smooth})}$ is a non-decreasing function which converges pointwise to $\delta^{(\text{hard})}$ on $(0,1) \cup (1,\infty)$.  
In the supplementary materials, we give sufficient conditions on $\delta_n^{(\text{smooth})}$ such that $\hat H_n^{(\text{smooth})} \Rightarrow H$. Consistency of 
\begin{eqnarray}
\hat{\gamma}_n^{(\text{smooth})} &:=& 
 \int_{[0,1]}  |2w - 1| \hat{H}_n^{(\text{smooth})}(dw) \nonumber 
      = \frac
      {\sum_{t=1}^n \delta_n^{(\text{smooth})} \left( \frac{\| \boldsymbol Z_t \|}{b(n/k)} \right) |2 W_t - 1|}
      {\sum_{t=1}^n \delta_n^{(\text{smooth})} \left( \frac{\| \boldsymbol Z_t \|}{b(n/k)} \right)}  \nonumber \\
  &=& \left( \sum_{t=1}^n \delta_n^{(\text{smooth})} \left( \frac{\| \boldsymbol Z_t \|}{b(n/k)} \right)  \right)^{-1}
    \sum_{t=1}^n \delta_n^{(\text{smooth})} \left( \frac{\| \boldsymbol Z_t \|}{b(n/k)} \right) 
	\frac{| Z_{t,1} - Z_{t,2} |}{| Z_{t,1} + Z_{t,2}| }  \label{eq:gammaHatConsist}
\end{eqnarray}
follows 
from the weak convergence of $\hat H_n^{(\text{smooth})}$ to $H$.

The conditions on $\delta_n^{(\text{smooth})}$ are related to its convergence rate. In the regular variation framework, points pile up near the origin at rate $n$. Thus, near the origin, $\delta_n^{(\text{smooth})}$ must converge to zero quickly enough to negate this effect. Away from the origin, $\delta_n^{(\text{smooth})}$ is allowed to converge to $\delta^{(\text{hard})}$ more slowly. 
In practice, since $n$ is fixed, the convergence rate of $\delta_n^{(\text{smooth})}$ is irrelevant. 
Typical of extremes procedures, in equations (\ref{eq:  smoothH})  and (\ref{eq:gammaHatConsist}) $b(n/k)$ is replaced by a suitably chosen threshold $r$.


\section{Procedure for Investigating Extreme Behavior} \label{sect3}

The focus of our project is not parameter estimation as discussed in Section \ref{sect2}.
We now develop the two parts of our method to find linear combinations of functions of covariates which are associated with extreme behavior.
We restrict our attention to linear combinations as we feel the model space would become unsearchable otherwise.

As the sample size is fixed in the remainder of this work, we omit the $n$ subscript from $\delta_n^{(\text{smooth})}$ and $\hat{\gamma}_n^{(\text{smooth})}$ hereafter.

\subsection{Optimizing Tail Dependence}
Assume for now that we work with a specific set of covariates.
We aim to find the linear combination which optimizes tail dependence between these covariates and the response variable in terms of $\hat \gamma^{(\text{smooth})}$.  
Let the response at time $t \in \mathbb{N}$ be given by the continuous random variable $Y_t$ (on its original scale), 
and let the $k-$dimensional random vector of continuous covariates be $\boldsymbol{X}_t = (X_{t,1},  \ldots , X_{t,k})^T$.  

To make use of the bivariate regular variation framework, we would like the response variable and the linear combination of covariates to have regularly varying marginal distributions.  
We can easily transform $Y_t$ to be approximately unit Fr\'echet.  
We define $Y_t^{**} = G^{-1}[\hat F_{Y}(Y_t)]$ where $\hat F_Y$ is an estimated stationary marginal distribution of $Y_t$ and $G$ is the unit Fr\'echet distribution function.

To ensure that our linear combination has a marginal which is approximately unit Fr\'echet, we do a two-step transformation procedure.  
We first transform each covariate to the $N(0,1)$ scale using a probability integral transformation, $X_{t,i}^* = \Phi^{-1} [\hat F_{X_i}(X_{t,i})]$, where $\Phi$ represents the Gaussian distribution function with mean 0 and variance 1.  
Define the vector $\boldsymbol{X}_t^* = (X_{t,1}^*, \ldots, X_{t,k}^*)^T$, letting $\Sigma^*$ denote its covariance matrix. 
We investigate functionals of the form 
\begin{equation} \label{eq:LINCOM}
\boldsymbol{X}_t^{*T} \boldsymbol{\beta}=\beta_1 X^*_{t,1} + \cdots + \beta_k X^*_{t,k},
\end{equation}
noting that $E[\boldsymbol{X}_t^{*T} \boldsymbol{\beta}] = 0$ and $	\text{Var}[\boldsymbol{X}_t^{*T} \boldsymbol{\beta}] = \boldsymbol{\beta}^T \Sigma^* \boldsymbol{\beta}$. 
For identifiability, we constrain $\boldsymbol{\beta}$ such that $\boldsymbol{\beta}^T \Sigma^* \boldsymbol{\beta} = 1$.

The second transformation ensures our function of the covariates is approximately unit Fr\'echet.
For optimization purposes, we assume that $\boldsymbol{X}_t^{*T}\boldsymbol{\beta}$ is approximately normal.
We then apply the transformation 
\begin{equation} \label{eq:transLinCom}
X_t^{**}(\boldsymbol{\beta}) = G^{-1}[\Phi(\boldsymbol{X}_t^{*T} \boldsymbol{\beta})].
\end{equation}
Employing the Gaussian cdf in (\ref{eq:transLinCom}) results in smooth behavior of the objective function in the optimization procedure.
If we were to use a rank-based transformation as we do for the response or in the first stage, the objective function would have a discontinuous jump at values of  $\boldsymbol{\beta}$ where the ordering of $\boldsymbol{X}_t^{*T}\boldsymbol{\beta}$ changes.

Define
$$
\hat{\boldsymbol{\beta}}^* = \underset{\{ \boldsymbol{\beta} \in \mathbb{R}^k : \boldsymbol{\beta}^T \hat \Sigma^* \boldsymbol{\beta} = 1 \}}{\operatorname{argmin}} \text{  } 
\frac
{\sum_{t = 1}^n \delta^{(\text{smooth})}( X_t^{**}(\boldsymbol{\beta}) + Y_t^{**}) \frac{|X_t^{**}(\boldsymbol{\beta}) - Y_t^{**} |}{X_t^{**}(\boldsymbol{\beta}) + Y_t^{**}} }
{\sum_{t = 1}^n \delta^{(\text{smooth})} ( X_t^{**}(\boldsymbol{\beta}) + Y_t^{**} )}.
$$
We use 
$\delta^{(\text{smooth})} (z ) 
= \Phi 
\left( 
  \frac{ z - r_0}{\sigma}
\right) 
$ 
as our weight function, where $\sigma$ determines the amount of smoothness and $r_0$ is a selected threshold.
Even with the implementation of the smoothed threshold, the optimization is non-trivial.  
Given $k$ covariates, the constraint $\boldsymbol{\beta}^T \Sigma^* \boldsymbol{\beta} = 1$ makes the optimization $k-1$ dimensional.
One can perform optimization directly on $\beta_1, \ldots, \beta_k$ using augmented Lagrangian methods available in the `alabama' package \citep{alabama} in R \citep{R}, but we found that this optimization required good starting values.
After transforming to standard polar coordinates: $\beta_1 = r \cos{\theta_1}$, 
$\ldots$, $\beta_{k-1} = r \left( \prod_{i=1}^{k-2} \sin{\theta_i} \right) \cos{\theta_{k-1}}$, $\beta_{k} = r \left( \prod_{i=1}^{k-1} \sin{\theta_i} \right)$, $r$ can be constrained in terms of $\bm \theta = (\theta_1, \ldots, \theta_{k-1})^T$ which has box constraints $\theta_j \in (0,\pi)$ for $j = 1, 2, \ldots, k-2$ and $\theta_{k-1} \in (0,2 \pi)$.  
We found that the differential evolution algorithm in `DEoptim' \citep{DEoptim} and the generalized simulated annealing algorithm in `GenSA' \citep{GenSA} performed equally well in optimizing $\bm \theta$.

\subsection{Model Comparison and Model Search} \label{SAdescr}
To compare the model fits corresponding to different covariate sets, we need a criterion which evaluates each model's level of tail dependence.
To protect against overfitting, we use a rather standard 10-fold cross-validation (CV) procedure; however, our CV score is based on the tail dependence metric $\gamma$ rather than a typical mean-squared-error metric, which is poorly suited for extremes.  
Given a particular set of covariates, we first 
obtain $Y_t^{**}$ and $X_{t,i}^*$ (for $i = 1,\ldots,k$) 
as described in Section 3.  
We then randomly partition these transformed observations into 10 equally sized subsets.  
Let the observation numbers corresponding to the $p^{th}$ partition be given by $\Gamma_p \subset \{1, 2, \ldots, n\}$.  
Similarly, let the indices corresponding to all observations except the $p^{th}$ partition be given by $\Gamma_{-p} = \{1, 2, \ldots, n\} \backslash \Gamma_p$.  
At the $p^{th}$ iteration (for $p = 1, \ldots, 10$), we obtain the parameter estimates using all observations except those in the $p^{th}$ partition:  
$$
\hat{\boldsymbol{\beta}}^{\left(-p\right)} = \underset{\{ \boldsymbol{\beta} \in \mathbb{R}^k: \boldsymbol{\beta}^T \hat{\Sigma}^* \boldsymbol{\beta} = 1\}}{\operatorname{argmin}} 
\frac
{\sum_{t \in \Gamma_{-p}} \delta^{(\text{smooth})} (X_t^{**}(\boldsymbol{\beta}) + Y_t^{**}) 
	\frac{|X_t^{**}(\boldsymbol{\beta}) - Y_t^{**}|}{X_t^{**}(\boldsymbol{\beta}) + Y_t^{**}}}
{\sum_{t \in \Gamma_{-p}} \delta^{(\text{smooth})}( X_t^{**}(\boldsymbol{\beta}) + Y_t^{**})}.
$$
After obtaining $\hat{\boldsymbol{\beta}}^{\left(-p\right)}$, we calculate $\hat{\gamma}_{\left(-p\right)}$ by applying $\hat{\boldsymbol{\beta}}^{\left(-p\right)}$ to the held out data:
$$
\hat{\gamma}_{\left(-p\right)} = 
\frac
{\sum_{t \in \Gamma_p} \delta^{(\text{smooth})} (X_t^{**}(\boldsymbol{\beta}) + Y_t^{**})
	\frac{|X_t^{**}(\hat{\boldsymbol{\beta}}^{\left(-i\right)}) - Y_t^{**}|}{X_t^{**}(\hat{\boldsymbol{\beta}}^{\left(-i\right)}) + Y_t^{**}}}
{ \sum_{t \in \Gamma_p} \delta^{(\text{smooth})} (X_t^{**}(\boldsymbol{\beta}) + Y_t^{**})}.
$$
After iterating over the 10 partitions, we calculate $CV = 10^{-1} \sum_{i = 1}^{10} \hat{\gamma}_{\left(-p\right)}$.  

Since it is impossible to fit and compare all potential models, we require a method to search the model space to identify good-fitting models.
To describe the model space, we consider binary strings $\boldsymbol{\omega}$ in the space $\{0,1\}^r$, where $r$ is the total number of covariates including interactions and transformations.  
In these strings, a 1 (or 0) in the $i^{th}$ position (for $i = 1,\ldots,r$) indicates the presence (or absence) of the $i^{th}$ covariate.  
Thus, $\boldsymbol{\omega}$ corresponds to the unique representation of one particular model.  
We denote the $CV$ value for the model represented by $\boldsymbol{\omega}$ with $CV(\boldsymbol{\omega})$, and we wish to find models for which $CV(\bm \omega)$ is small.
Thus, we frame variable selection as a combinatorial optimization problem. To search the model space we employ simulated annealing, which is known to give good solutions in these types of problems \citep{kirkpatrick1983}.

We implement a slightly modified version of the simulated annealing procedure utilized in the R optim function (using the SANN method).  
We begin with an initial value, $\boldsymbol{\omega}_0 \in \{0,1\}^r$ and rely on a function $f: \{0,1\}^r \rightarrow \{0,1\}^r$ to choose a new string.  
We can construct $f$ to exclude certain undesired models, such as models with highly correlated covariates. We describe the function $f$ we use in our analysis in the Supplementary Materials. 

At the $j^{th}$ step we compare $CV(f(\boldsymbol{\omega}_{j-1}))$ to $CV(\boldsymbol{\omega}_{j-1})$.  If $CV(\boldsymbol{\omega}_{j-1}) > CV(f(\boldsymbol{\omega}_{j-1}))$ then we define $\boldsymbol{\omega}_j := f(\boldsymbol{\omega}_{j-1})$ and proceed to the next iteration.  If $CV(\boldsymbol{\omega}_{j-1}) \leq CV(f(\boldsymbol{\omega}_{j-1}))$ then 
$$
\boldsymbol{\omega}_j := \begin{cases}
f(\boldsymbol{\omega}_{j-1}) &\mbox{with probability    }~ \exp\{-\Delta CV_j/\text{Temp}_j\} \\
\boldsymbol{\omega}_{j-1} &\mbox{with probability    }~ 1-\exp\{-\Delta CV_j/\text{Temp}_j\} \end{cases}
$$
where Temp$_j$ is the current global temperature in the simulated annealing process and $\Delta CV_j = CV(f(\boldsymbol{\omega}_{j-1}))-CV(\boldsymbol{\omega}_{j-1})$.  The global temperature is a parameter that is lowered throughout the optimization according to a cooling schedule.  As in the R function optim using the SANN method, we use the logarithmic cooling schedule outlined in \cite{belisle1992}.  When the global temperature is high the process is more likely to move to $\boldsymbol{\omega}$s with higher $CV$ values, reducing the chances of finding a local optimum.  When the global temperature is low, the process is unlikely to move to $\boldsymbol{\omega}$s with higher $CV$ values. 
Atypical for simulated annealing, our goal is not necessarily to find the unique global optimum, but to identify several models with very good scores and investigate their commonalities and differences.


\section{Simulation Study}
\label{sec4}
\subsection{Description of Simulated Data}
We randomly generate 5,000 independent realizations of five covariates, $X_1, X_2, X_3, X_4, X_5$.  
The first four covariates are four-dimensional Gaussian with non-identity covariance matrix.  
The fifth covariate is drawn independently of the first four covariates uniformly on the unit interval, i.e. $X_5 \sim U(0,1)$.  
The response, $Y$, is a linear combination of functions of the five covariates plus noise: 
\begin{equation}
Y_t = -.3 X_{t,1} + X_{t,2} -.75 X_{t,4} - (X_{t,2})^2 + 6 \Phi[(X_{t,1}-X_{1;.95})/.35] X_{t,5} + \varepsilon_t, \label{eq:SIM}
\end{equation}
where $X_{1;.95}$ represents the .95 quantile of $X_1$ and $\varepsilon_t$ $\sim$ iid N(0, .00125).
The idea is to create a response which has a nonlinear relationship with the covariates which only appears when conditions are extreme.  
Specifically, the term $6 \Phi[(X_{t,1}-X_{1;.95})/.35] X_{t,5}$ only contributes to $Y_t$ when $X_{t,1}$ is extremely large.

\begin{figure}[h]
\begin{center}
\includegraphics[width=0.9\textwidth]{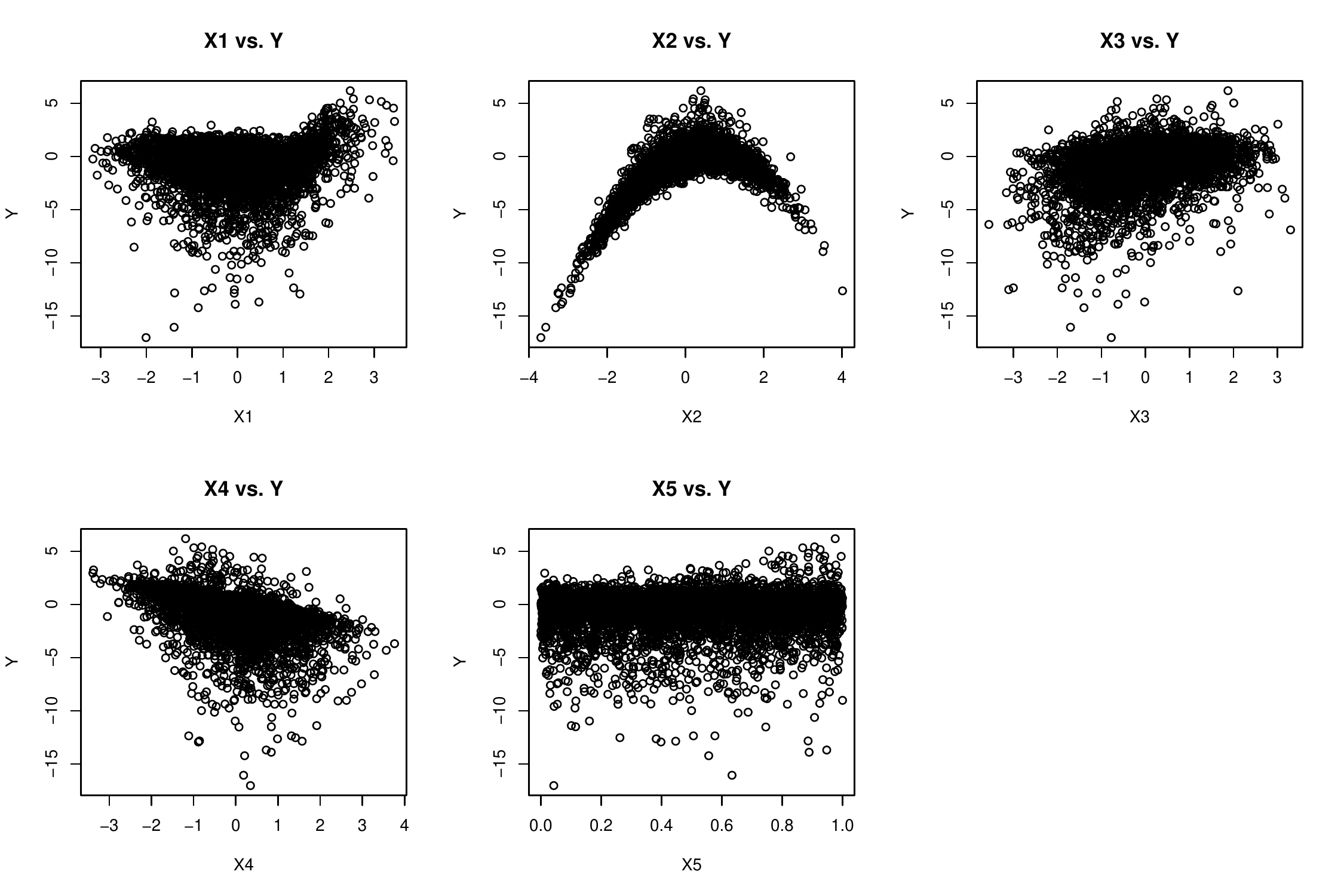}
\caption{Scatterplots for each of the five covariates versus the response in the simulation study.  All variables are on their original scales.}
\label{fig:simulation_scatterplots}
\end{center}
\end{figure}

Figure~~\ref{fig:simulation_scatterplots} shows scatterplots of the response variable versus each of the five covariates; 
we focus on the relationships driving the large response values. 
Large values of $Y_t$ are clearly associated with large values of $X_{t,1}$.  
The quadratic term causes large values of the response to occur when $X_{t,2}$ is between 0 and 1.5.
$X_{t,3}$ does not seem to be related to large response values,  and  $X_{t,4}$'s relationship with the response appears roughly linear.
Finally, the evidence of $X_{t,5}$'s influence on the extreme behavior is slight, and its interaction with $X_{t,1}$ is not apparent from these plots.

\subsection{Model Selection and Estimation}
\label{sec: modelSel}
To test our model selection procedure, we consider the models M1-M6 listed below.  
The first model includes all five covariates.  
The models M2-M4 each leave out a single covariate:  $X_3, X_2,$ $X_1$ respectively. 
Model five leaves out $X_3$ but adds an interaction between $X_1$ and $X_5$.  
Model six adds $(X_2)^2$ to model five.  
Note that none of these models corresponds exactly to equation~(\ref{eq:SIM}), but that M6 is closest that it includes some type of interaction between $X_1$ and $X_5$ and the quadratic behavior of $X_2$.

\noindent 
M1: $X_t^{**} = [X^*_{t,1}, X^*_{t,2}, X^*_{t,3}, X^*_{t,4}, X^*_{t,5}] \boldsymbol{\beta}_1$\\
M2: $X_t^{**} = [X^*_{t,1}, X^*_{t,2}, X^*_{t,4}, X^*_{t,5}] \boldsymbol{\beta}_2$\\
M3: $X_t^{**} = [X^*_{t,1}, X^*_{t,3}, X^*_{t,4}, X^*_{t,5}] \boldsymbol{\beta}_3$\\
M4: $X_t^{**} = [X^*_{t,2}, X^*_{t,3}, X^*_{t,4}, X^*_{t,5}] \boldsymbol{\beta}_4$\\
M5: $X_t^{**} = [X^*_{t,1}, X^*_{t,2}, X^*_{t,4}, X^*_{t,5}, X^*_{t,1} \times X_{t,5}] \boldsymbol{\beta}_5$\\
M6: $X_t^{**} = [X^*_{t,1}, X^*_{t,2}, X^*_{t,4}, X^*_{t,5}, X^*_{t,1} \times X^*_{t,5}, (X^*_{t,2})^2] \boldsymbol{\beta}_6$\\

We optimize the tail dependence with a smooth threshold where $\sigma = 1.25$ and $r_0 = 40$, the .95 quantile of the radial components $\| \bm z_t \|$.  
Marginal transformations from the original scale were based on the rank transform.
Table \ref{tab:SimModComp} gives $\hat \gamma$ and the CV value for each of the six models.  
Comparing models M1 and M2 shows that the CV method is effective in protecting against overfitting.
The overall score $\hat \gamma$ is higher (worse) for M2 as it is a submodel of M1; however, the CV score for M2 is better indicating that  $X_{t,3}$ is not useful for describing extreme response values.
The CV score for M4 shows that leaving out $X_1$ is clearly not a good idea as the CV value is by far the highest of the six models.  
M5, which adds the interaction of $X_1$ and $X_5$, gives a noticeably improved CV value, and M6, which is closest to the generating model has the best CV score.

%

\begin{table}[ht]
\centering
\caption{The score (the optimized value of $\hat{\gamma}$) and the 10-fold cross-validation value for each of the six models considered in the simulation study.}
\begin{tabular}{| l | c | c | c | c | c | c |}
  \hline
 & M1 & M2 & M3 & M4 & M5 & M6 \\ 
  \hline
$\hat{\gamma}$ & 0.4930 & 0.4950 & 0.5125 & 0.6053 & 0.4603 & 0.4060 \\ 
$CV$ & 0.5052 & 0.5017 & 0.5240 & 0.6196 & 0.4703 & 0.4120 \\ 
   \hline
\end{tabular}
\label{tab:SimModComp}
\end{table}

\begin{figure}[h]
\begin{center}
\includegraphics[width=0.9\textwidth]{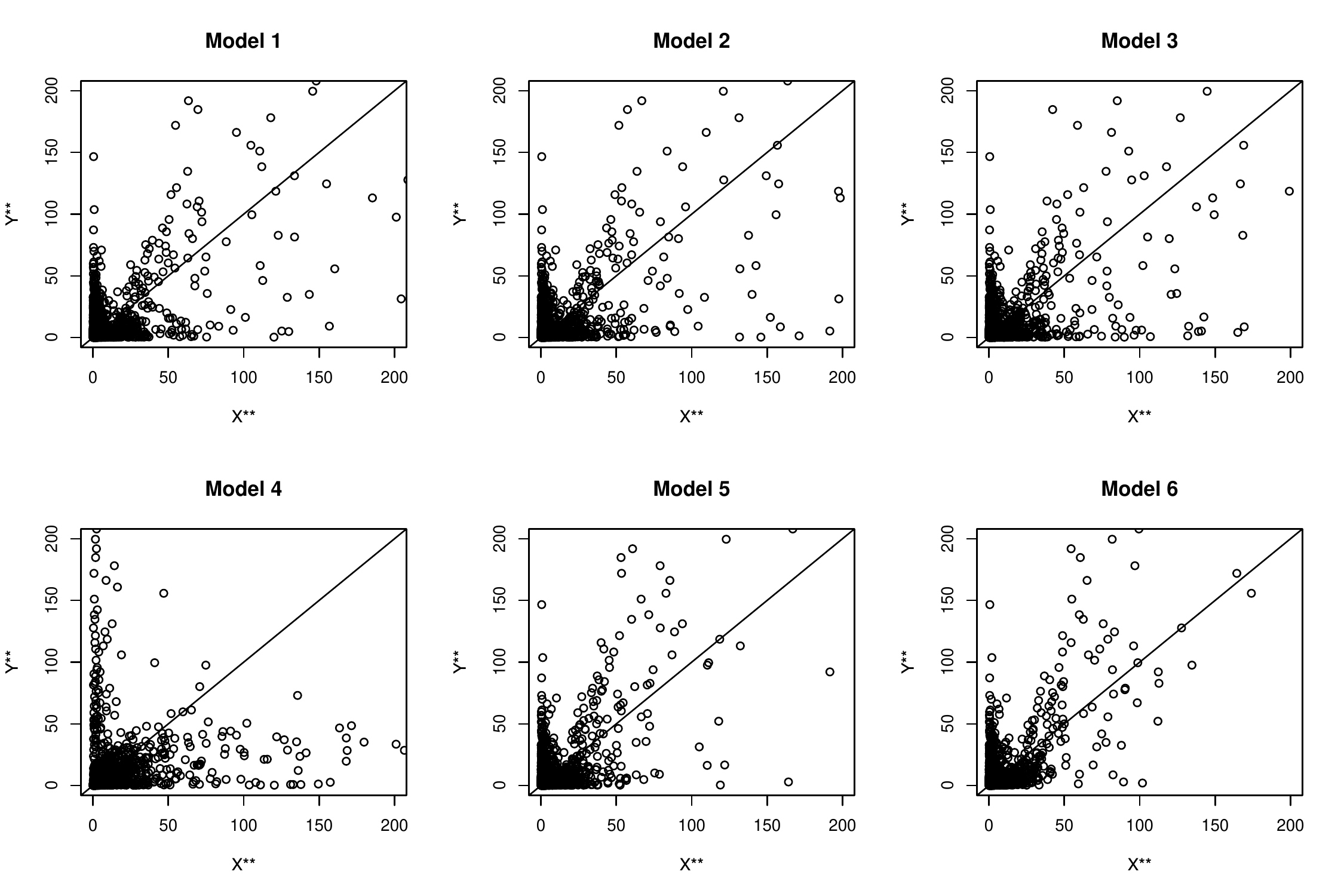}
\caption{For each of the six models we consider in the simulation study, the scatterplot of $x_t^{**}(\hat{\boldsymbol{\beta}}^*)$ versus $y_t^{**}$ is given.  Models that yield a linear combination with a higher degree of tail dependence with the response will result in a scatterplot with large points closer to the identity line. Note the figures have been zoomed in to show the $[0,200] \times [0,200]$ box to better illustrate the difference in behavior, but this viewing window does exclude the largest values.  Expanding the plotting range does not affect the qualitative behavior.}
\label{fig:unitFrechet_scatterplots}
\end{center}
\end{figure}

Figure~\ref{fig:unitFrechet_scatterplots} gives a visual way to assess tail dependence associated with each model.  
For each model, $x_t^{**}(\hat{\boldsymbol{\beta}}^*)$ is plotted versus $y_t^{**}$, where we use lower case to denote the realization from (\ref{eq:SIM}).  Note that the large points for the models with lower $\hat{\gamma}$ scores and CV values, like M5 and M6, occur in the interior of the positive orthant, whereas models with low $\hat{\gamma}$s (M4) have more points near the axes.

Table~\ref{tab:M6ests} reports the parameter estimates with nonparametric bootstrap \citep{efron1993} standard errors for M6.  
Due to the constraint and the marginal transformations, only the relative magnitude and sign of the parameter estimates are interpretable.
The estimate with largest magnitude corresponds to $X_1$, indicating that it plays an important role in describing extreme levels of the response.  
Standard errors indicate that the $\hat \beta_{X_1}$, $\hat \beta_{X_4}$, $\hat \beta_{X_1X_5}$ and $\hat \beta_{(X_2)^2}$ terms are significant, and the signs of these agree with the behavior shown in Figure \ref{fig:simulation_scatterplots}.
The fact that $\hat \beta_{X_5}$ is not significant is also sensible, as $X_5$'s only contribution in the generating equation (\ref{eq:SIM}) is via the interaction with $X_1$.


\begin{table}[h]
\centering
\caption{The parameter estimates for M6 in the simulation study with bootstrap standard errors in parentheses.}
\begin{tabular}{| l | c | c | c | c | c | c |}
\hline
Coefficient & $\hat{\beta}_{X_1}$ & $\hat{\beta}_{X_2}$ & $\hat{\beta}_{X_4}$ & $\hat{\beta}_{X_5}$ & $\hat{\beta}_{X_1 X_5}$ & $\hat{\beta}_{(X_2)^2}$  \\ \hline
Estimate & .61 (.12) & .10 (.10) & -.22 (.07) & -.19 (.14) & .47 (.15) & -.17 (.08) \\ \hline
\end{tabular}
\label{tab:M6ests}
\end{table}

\subsection{Data Mining Procedure}
In order to evaluate our automated model search technique outlined in Section~\ref{SAdescr}, we expand the simulation study. In addition to the seven covariates used in M1-M6, we randomly generate 100 independent $N(0,1)$ random vectors. We utilize the model search procedure based on simulated annealing using the full set of 107 variables. 

We start eight optimizations in parallel, using a randomly generated starting value for each. 
Since each string is given a set amount of time to search the model space, the algorithm will likely not reach the global optimum. 
Rather, we aim to determine if the good scoring models capture the known causes for extreme responses for this simulated data.

Table~\ref{tab:SAtest} gives the strings with the best $CV$ scores in the simulated annealing based automated model search procedure. 
M6, from Section \ref{sec: modelSel} has a $CV$ score of .4120. 
We note that the top three strings from the model search have $CV$ scores that are close to this value. 
More importantly, these best scoring models clearly convey the driving factors for extreme behavior, as all include $X_1$, $X_4$, a quadratic term for $X_2$, and the $X_1 \times X_5$ interaction.


\begin{table}[h]
\centering
\caption{Cross validation scores and selected covariates of the best three models found by the automated model search algorithm.}
\begin{tabular}{| r | r | l |}
\hline
Model	& $CV$ & Variables  \\ \hline
M6 & .4120 & $X^*_{t,1}, X^*_{t,2}, X^*_{t,4}, X^*_{t,5}, X^*_{t,1} \times X^*_{t,5}, (X^*_{t,2})^2$  \\  \hline \hline
SA String 1 & .4253 & $X^*_{t,1}, X^*_{t,4}, X^*_{t,1} \times X^*_{t,5}, (X^*_{t,2})^2, X^*_{t,15}, X^*_{t,98}$  \\  \hline
SA String 2 & .4264 & $X^*_{t,1}, X^*_{t,4}, X^*_{t,1} \times X^*_{t,5}, (X^*_{t,2})^2, X^*_{t,66}, X^*_{t,79}$  \\  \hline
SA String 6 & .4271 & $X^*_{t,1}, X^*_{t,4}, X^*_{t,1} \times X^*_{t,5}, (X^*_{t,2})^2, X^*_{t,9}, X^*_{t,84}$  \\  \hline
\end{tabular}
\label{tab:SAtest}
\end{table}

Also within the simulation study, we assessed the algorithm's sensitivity to the threshold smoothness, but found that doubling or halving the value of $\sigma$ had very little effect on parameter estimates.
We also compare our results to other possible approaches, but note that none of the other approaches were designed to optimize tail dependence.
We outperform regression approaches and extremes regression approaches in terms of Kendall's $\tau$ applied to the joint tail.
Details on both the sensitivity analysis and comparison to other methods can be found in the supplementary materials.

\section{Application to Ground Level Ozone Pollution} \label{sect4}
  


We now employ our method in a data mining capacity in order to better understand the meteorological drivers of extreme ground level ozone.  We analyze ozone data for Atlanta, Georgia and Charlotte, North Carolina because of their geographical proximity and consistent data records for ozone.


\subsection{Data}
An EPA website\footnote{See http://www.epa.gov/airdata/ad\_maps.html.} provides ground level ozone data as well as data on other pollutants.
We selected station 13-121-0055 in Atlanta and station 37-119-1005 in Charlotte because of their long data records and relative lack of missing values.
Although the ozone level is measured hourly at each station, the response variable we use is the maximum eight hour average ozone as it is a value on which United States National Ambient Air Quality Standards are based.  
The EPA-defined ozone season for North Carolina is April through October while the ozone season for Georgia is March through October.  
In our analysis, we use daily responses from April through October for both locations.  
Our analysis is based on data from the years 1992 through 2010, providing a total of 4,037 observations for Atlanta and 4,055 observations for Charlotte.

Ground level ozone readings have been decreasing over most of the United States in recent years.  
On its website, the EPA reports that ``nationally, average ozone levels declined in the 1980s, leveled off in the 1990s, and showed a notable decline after 2002.''\footnote{See http://www.epa.gov/airtrends/ozone.html.}  
This is a trend that we notice in our exploratory data analysis.  
To account for non-stationarity, we transform the response variable by partitioning the data into nonoverlapping four-year blocks.  
In each block, we fit a gamma distribution to the observations below the .95 quantile and a generalized Pareto distribution for observations above the .95 quantile.  
We then use these estimated distribution functions to transform the response to unit Fr\'echet via a probability integral transformation.  
We employ a parametric model for marginal transformation as a rank transform resulted in common values in the tail if blocks have the same number of observations.

Not surprisingly, a seasonal effect is also apparent in the ozone data. 
Because our aim is to link extreme ozone levels to meteorological conditions, we do not deseasonalize the ozone data.  
Rather, we assume that the seasonal response is well captured by conditioning on meteorological variables which themselves exhibit seasonality.

We obtain meteorological covariates from the North American Regional Reanalysis (NARR)\footnote{NARR data provided by the NOAA/OAR/ESRL PSD, Boulder, Colorado, USA, from their Web site at http://www.esrl.noaa.gov/psd/} \citep{mesinger2006}.  
The data are in gridded cells, approximately 30km by 30km in size, differing from our ozone data which correspond to point locations.  Understanding the relationship between ozone and meteorological variables of large spatial scales is motivated by the larger project's goal of investigating the simulation of air quality extremes in atmospheric chemistry models.   
The NARR provides a large number of meteorological variables.  
Based on guidance from the collaborating atmospheric chemists, we initially use 18 NARR covariates in our models; however, we also consider these variables on different spatial scales and transformations of these variables leading to a longer overall list of possible covariates.  
We expect that many of these variables will be irrelevant in terms of explaining extreme ozone behavior.  
Furthermore, we aim to explore whether any interactions between covariates are useful to explain extreme ozone behavior.  
Hence, covariate selection is a primary goal of this study.

We do not include information about the ozone precursors $\text{NO}_{\text{x}}$ and VOCs among our covariates.
The EPA monitors NO and NO$_2$, and although the $\text{NO}_{\text{x}}$ data record is not as extensive as it is for ozone, many studies (e.g., \cite{eastoe2009}) have found $\text{NO}_{\text{x}}$ measurements to be helpful when modeling ozone.
The larger aim of our project is to provide information to improve atmospheric chemistry models which require information about $\text{NO}_{\text{x}}$ {\em emissions}, rather than measurements.
$\text{NO}_{\text{x}}$ emissions are not known at the daily level, and are presumed by modelers to be relatively constant.
Our aim is to link extreme ozone to meteorological conditions, and we believe that the daily variability found in $\text{NO}_{\text{x}}$ measurements is largely attributable to meteorology rather than fluctuations in emissions.

\subsubsection{Handling a Semi-continuous Covariate:  Precipitation}
The method described in Section 3 requires continuous covariates to perform the two-step marginal transformation leading to $X^{**}_t(\bm \beta)$.
We wish to investigate precipitation's effect on extreme ozone, but precipitation has a positive probability of being exactly zero.
Exploratory analysis indicates that the presence of precipitation likely affects extreme ozone, but the amount of precipitation may not be important.  Thus, we extend the method to account for variables like precipitation by including a precipitation indicator and interactions with this indicator.

In Section 3, it was critical that the distribution of ${\bm X^*_t}^T \bm \beta$ be known for any $\bm \beta$, which we achieved with the constraint $\boldsymbol{\beta}^T \hat{\Sigma}^* \boldsymbol{\beta} = 1$.
Including a precipitation indicator complicates matters, but we model in a way such that ${\bm X^*_t}^T \bm \beta$ will have a known distribution which is a mixture of Gaussians.

Let $X_{t,P}$ be the amount of precipitation at time $t$, where $P(X_{t,P}=0)>0$.  
Assume there are $k+l-m$ total covariates included:  $k$ `main effects' as before, and $l$ effects to be included in precipitation interaction terms, $m$ of which were already included as main effects.
Including the precipitation covariate and interactions  and letting $X_{t,1}, \ldots, X_{t,m}$ be the overlapping covariates changes equation~(\ref{eq:LINCOM}) to
\begin{eqnarray}
 \label{eq:RHS}
\boldsymbol{X}_t^{*T} \boldsymbol{\beta} &=& \beta_1 X^*_{t,1} + \cdots + \beta_m X^{*}
_{t,m} + \beta_{m+1} X^*_{t,m+1} + \cdots + \beta_k X^*_{t,k} +   \\
& &\mathbb{I}_{\{X_{t,P} > c\}} ( \beta_0^{(P)}  + \beta_1^{(P)} X^*_{t,1} + \cdots + \beta_m^{(P)} X^*_{t,m} + \beta_{m+1}^{(P)} X^*_{t,k+1} + \cdots + \beta_{l}^{(P)} X^*_{t,k+l-m} ), \nonumber
\end{eqnarray}
for $\boldsymbol{\beta} =  (\boldsymbol{\beta}_{(P^C)}^T,\boldsymbol{\beta}_{(P)}^T)^T = ((\beta_1,\ldots,\beta_k),(\beta_0^{(P)},\beta_1^{(P)},\ldots,\beta_l^{(P)}))^T$.  
As before, if $\boldsymbol{X}_t^{*T} \boldsymbol{\beta}|(X_{t,P} \leq c) \sim N(0,1)$ under the constraint $\boldsymbol{\beta}_{(P^C)}^T \Sigma^* \boldsymbol{\beta}_{(P^C)}=1$.  
However, given $X_{t,P} > c$,~(\ref{eq:RHS}) becomes
\begin{eqnarray}
\boldsymbol{X}_t^{*T} \boldsymbol{\beta}|(X_{t,P} > c) &=& (\beta_1+\beta_1^{(P)}) X^*_{t,1} + \cdots + (\beta_m+\beta_m^{(P)}) X^*_{t,m} + \nonumber \\
 & & \beta_{m+1} X^*_{t,m+1} + \cdots + \beta_k X^*_{t,k} + \beta_0^{(P)} + \nonumber \\
 & & \beta_{m+1}^{(P)} X^*_{t,k+1} + \cdots + \beta_{l}^{(P)} X^*_{t,k+l-m}, \label{eq:COND}
\end{eqnarray}  
which is distributed $N(\beta_0^{(P)},\boldsymbol{\lambda}^T \Psi^* \boldsymbol{\lambda})$ where $\Psi^*$ is the covariance matrix of the continuous covariates and 
$\boldsymbol{\lambda} =((\beta_1+\beta_1^{(P)}),\ldots,(\beta_m+\beta_m^{(P)}),\beta_{m+1},\ldots,\beta_k,\beta_{m+1}^{(P)},\ldots,\beta_{l}^{(P)})^T.$
Since $ \boldsymbol{X}_t^{*T} \boldsymbol{\beta}$ has a known distribution for any $\bm \beta$, we can proceed with the second transformation as before, employing sample covariance matrices and the observed mixture proportion.

\subsection{Data Mining Procedure and Results} \label{sect Data Mining}

As we are not able to fit all possible models, our data mining approach evolved as a multiple-step procedure, and we will discuss results at each step.
We use the Yellowstone computing system \citep{CISL2012} to perform computations.  Also, based on exploratory analysis, we employ a smooth threshold with mean equal to the .95 quantile of the radial components and $\sigma=1.25$.

\subsubsection{Model Exploration: Four Variable Models}
As a first step, we fit all possible models with up to four covariates because an exhaustive search of such models is possible.
There are nearly 35,000 models of this type, and we fit all of the approximately 10,000 models that do not include highly correlated covariates.  
We allow the models to include the precipitation indicator and interactions between continuous covariates and the precipitation indicator, but do not consider interactions between continuous covariates.  
We define the precipitation indicator to be $\mathbb{I}{\{X_{t,P} > .01\text{in.}\}}$.


Table~\ref{tab:BestFourVar} reports the covariates in the five models with the lowest CV scores for Atlanta and Charlotte.
In Atlanta, variables such as temperature, wind speed,  downward shortwave radiative flux (dswrf, a measure of sunshine), the precipitation indicator, height of the planetary boundary layer (hpbl) at different times of day, and relative humidity seem to be in the best fitting models.
In Charlotte, similar variables appear along with northwest or west (shown as a negative coefficient on east) wind directions.
Most of these variables are not surprising, as one would suspect high ozone to be associated with hot sunny days with low wind speeds.
We view the results from this first step mostly as confirmatory:  the approach is choosing sensible covariates when limited to only four variables at a time.
That the precipitation indicator appears is somewhat interesting as \cite{jacob2009} noted that precipitation has little effect on ground level ozone pollution.
While precipitation may have little effect on mean levels of ground level ozone, it seems reasonable that the most extreme ozone levels do not occur on days where there is precipitation.

We also note that the four-covariate models with the best CV scores are identical for Atlanta and Charlotte, and this allows us to compare the parameter estimates for these common models.
Table~\ref{tab:BestFourVar} also gives the parameter estimates and bootstrap standard errors (based on 640 bootstrap replications) for the top model.
We note that the corresponding parameter estimates are similar and the signs of the parameter estimates are sensible.
 Both locations show that air temperature and downward shortwave radiation flux have a positive relationship with extreme ozone while wind speed and precipitation have negative relationships.  
Standard errors show there is large uncertainty associated with the parameter estimates, but our CV-based model selection procedure should protect us from identifying irrelevant covariates.


It is clear that since we are able to include only four variables at a time, we have limited ability to explore the model space.  
However, results from this first step suggest a method for further exploration of the model space.


\begin{table}[h]
\centering
\caption{Covariates for the top five models containing four covariates for Atlanta and Charlotte are given along with their respective CV scores.  
Parameter estimates with bootstrap standard errors (based on 640 bootstrap replications) are reported for the top model at each location.}
\begin{tabular}{| l | l | l | l | l | l |}
\hline
Model & CV & & & & \\
\hline
Atlanta 1 & .5398 & temp & wnd spd & dswrf & precip   \\ 
  & & .45 (.25) & -.44 (.36) & .53 (.64) & -5.82 (.20) \\ \hline
Atlanta 2 & .5508 & temp & wnd spd & rel hum & precip 		 \\ 
  & & .59 & -.46  & -.28 & -5.46     \\ \hline
Atlanta 3 & .5512 & temp & wnd spd & dswrf & cape 			 \\ 
  & & .80 & -.39  & .34 & -.35     \\ \hline
Atlanta 4 & .5513 & temp & wnd spd & hpbl 7am & cape 		 \\ 
  & & .87 & -.36  & -.24 & -.28     \\ \hline
Atlanta 5 & .5519 & temp & hpbl 7am & rel hum & precip 		 \\ 
  & & .53 & -.39  & -.48 & -5.10     \\ \hline \hline
Charlotte 1 & .5452 & temp & wnd spd & dswrf & precip  \\ 
  & & .44 (.33) & -.50 (.42) & .53 (.61) & -5.86 (.43) \\ \hline
Charlotte 2 & .5770 & temp & wnd spd & NW wind & dswrf \\ 
  & & .54 & -.50  & .12 & .38     \\ \hline
Charlotte 3 & .5772 & temp & wnd spd & dswrf & rel hum \\ 
  & & .52 & -.44  & .34 & -.19     \\ \hline
Charlotte 4 & .5774 & temp & wnd spd & E wind & dswrf \\ 
  & & .54 & -.46  & -.10 & .45     \\ \hline
Charlotte 5 & .5781 & temp & wnd spd & dswrf & tcdc \\ 
  & & .55 & -.51  & .46 & .15     \\ \hline
\end{tabular}
\label{tab:BestFourVar}
\end{table}



\subsubsection{`Core Plus Four' Model Search}
The main result from the first model search stage is that high temperature, high sunshine, low wind speed, and lack of precipitation seem to provide good conditions for extreme ozone events. 
These conditions are largely explained by the four `core' variables which appear in the best fitting models for Atlanta and Charlotte: temperature, wind speed, dswrf, and the precipitation indicator.
We continue our model search by fitting all possible models which include these four variables, plus up to four additional main effects.
A total of 534 models were fit at this stage.



Results from the first stage also suggested slightly altering our list of covariates.
One change that was made was to include only the minimum and maximum hpbl values rather than the 8 
values recorded throughout each day.


Including dswrf as a core variable led us to change how we dealt with several other cloud variables which we found to be strongly correlated with dswrf.
We define a new variable which is residual to the information in dswrf.
The new variable is a linear combination of five of the cloud variables in the NARR: $\boldsymbol{x} = [x_{\text{cdcon}}, x_{\text{cdlyr}}, x_{\text{lcdc}}, x_{\text{mcdc}}, x_{\text{hcdc}}]^T$. 
Specifically, we find the parameter vector of unit length, $\boldsymbol{a}$
such that $	\text{Var}(\boldsymbol{a}^T \boldsymbol{x}) = \boldsymbol{a}^T \Sigma \boldsymbol{a}$ is maximized and $Cov(x_{\text{dswrf}},\boldsymbol{a}^T \boldsymbol{x}) = 0$.
We estimate $\boldsymbol{a}$ via constrained optimization at several locations in the East and Southeast United States (including Atlanta and Charlotte) and find $\boldsymbol{a}$ to be similar at all locations.  
Thus, we define the new cloud variable: 
$$x_{\text{new.cloud}} = .47x_{\text{cdcon}} - .45x_{\text{cdlyr}} - .37x_{\text{lcdc}} + .46x_{\text{mcdc}} + .48x_{\text{hcdc}}.$$ 
This new cloud variable can be loosely interpreted as a contrast between high level clouds (those with positive coefficients) and low level clouds (those with negative).

Table~\ref{tab:coreplusfouratlchar} compares the top five `core-plus-four' models at Atlanta and Charlotte to the core-only model.
In Atlanta, we see a convincing drop in the CV scores of the best fitting core-plus-four models compared to the core-only model.
The top models tend to include minimum planetary boundary layer height (in 5 of the top 5 and 9 of the top 10), relative humidity (5/5 and 9/10), tropospheric height (3/5 and 5/10) and NE wind direction (3/5 and 4/10).
The negative coefficients indicate that lower levels of planetary boundary layer height and relative humidity tend to be associated with extreme ozone, and the negative coefficient of the NE wind direction would indicate that extreme ozone tends to occur in Atlanta when wind is from the southwest.

In Charlotte, we see a less convincing drop in the CV scores when we compare the best fitting core-plus-four models to the core only model.
However, there are some variables which are associated with most of the best fitting models.
The top models in Charlotte tend to include the new cloud variable (in 3 of the top 5 and 7 of the top 10), tropospheric height (3/5 and 6/10), and E wind direction (3/5 and 6/10).
Interestingly, hpbl which seems to have a clare signal with extreme ozone in Atlanta, appears in few of the best fitting Charlotte models.
The differences in the two cities' variables may illustrate that different factors lead to extreme ozone in the two cities, and the difference in the predominant wind direction may illustrate local differences in emissions sources.
That the tropospheric height variable has a negative coefficient in Charlotte and a positive coefficient in Atlanta illustrates some of the difficulty in interpreting the parameter estimates.
The difference in sign between the two cities may be due to the fact that in Atlanta, tropospheric height appears in models which also include planetary boundary layer height, whereas this was not the case in Charlotte.

\begin{table}[ht]
\begin{center}
\caption{CV scores for the core only model and the top five `core-plus-four' models for both Atlanta and Charlotte. We also include the parameter estimates for the four non-core covariates.}
\begin{tabular}{lrrrrr}
  \hline
Rank  & CV & \multicolumn{4}{c}{Covariates Added to Core Model}\\
  \hline
Atl Core  & 0.5398 &  & & & \\
  \hline
Atl 1 & 0.5046 &  hpbl min & rel hum & pres & ht tropo \\
  &        &            -0.34     & -0.19& -0.12& 0.22 \\
  \hline
Atl 2 & 0.5075 &  hpbl min & rel hum & NE wnd & ht tropo \\
  &        &            -0.30     & -0.17& -0.16   & 0.15 \\
  \hline
Atl 3 & 0.5083 &  hpbl max & hpbl min & rel hum & ht tropo \\
  &        &            -0.10     &-0.32      & -0.35&  0.16 \\
  \hline
Atl 4 & 0.5090 &  hpbl min & rel hum & NE wnd & pres \\
  &        &            -0.39     & -0.10& -0.19   &-0.06 \\
  \hline
Atl 5 & 0.5097 &  hpbl min & rel hum & NE wnd & lwrf \\
  &        &            -0.36     & -0.17& -0.17   & 0.08 \\
  \hline
    \hline
Char Core & 0.5452 &   &  &  & \\
  \hline
Char 1 & 0.5412 &  E wnd & pres & lwrf & ht tropo \\
  &        &            -0.11  & 0.08 & 0.13  & -0.12 \\
  \hline
Char 2 & 0.5415 &  cloud & E wnd & pres chng & pres \\
  &        &            -0.11      & -0.09  & -0.04       & 0.14 \\
  \hline
Char 3 & 0.5415 &  cloud & E wnd & ht tropo &  \\
  &        &            -0.14      & -0.10  & -0.11 &  \\
  \hline
Char 4 & 0.5420 &  hpbl max & NW wnd & lwrf & ht tropo \\
  &        &            -0.10     & 0.06    & 0.04  &-0.05 \\
  \hline
Char 5 & 0.5421 &  cloud & rel hum & N wnd & pres chng \\
  &        &            -0.06      & -0.21&  0.04  &-0.07  \\
  \hline
\end{tabular}
\label{tab:coreplusfouratlchar}
\end{center}
\end{table}

Table \ref{tab:coreplusfouratlcharBS} gives bootstrapped standard errors for the best fitting models in Atlanta and Charlotte.
Because we are using a small subset of extreme data, and because these models include a large number of covariates which are likely dependent, it is not surprising that the standard errors are quite large.
Because our aim is to uncover possible covariates for further exploration rather than to give a definitive model, we are not overly concerned with the large standard errors.

\begin{table}[ht]
\begin{center}
\caption{Parameter estimates of the best `core plus four' models for Atlanta and Charlotte with bootstrap standard errors in parentheses.}
\begin{tabular}{lrrrr}
  \hline
Atlanta & temp & wnd spd & dswrf & precip \\ 
 &  .40 (.27)&  -.31 (.29)&  .29 (.50)&  -2.12 (.96) \\
 &  hpbl min & rel hum & pres & ht ropo  \\
 & -.34 (.47) &  -.19 (.28)&  -.12 (.26) &  .22 (.28) \\ \hline \hline
Charlotte & temp & wnd spd & dswrf & precip \\
 & .46 (.31) & -.41 (.34) & .55 (.48) & -4.35 (.89) \\
 & E wnd & pres & lwrf & ht tropo   \\
 & -.11 (.39) & .08 (.17) &  .13 (.24) & -.12 (.27) \\ \hline
\end{tabular}
\label{tab:coreplusfouratlcharBS}
\end{center}
\end{table}

\subsubsection{Automated Model Search Procedure}
Our model search procedure thus far has been limited to at most eight main effects.
We would like to further explore the model space to investigate whether interactions or a larger number of covariates would show even stronger tail dependence.
Because a systematic model search becomes infeasible, we perform an automated model search utilizing our simulated annealing procedure described in Section \ref{SAdescr}.

Possible covariates include all the main effects considered in the previous `core plus four' exploration, 77 interactions between continuous covariates, and 15 interactions between continuous covariates and the precipitation indicator. 
We include the four core variables in all considered models, as this reduces the search to a region of the model space where extremes are known to occur. 
Starting values are chosen by using the best core plus four models at each location. 
We perform 640 runs at each location. The simulated annealing procedure requires a choice for the function $f$. We describe our choice in the Supplementary Materials.

The covariates in the top five models at each location are given in 
Table~\ref{tab:SAmodelsAtlChar}.  
In Atlanta, and to a greater extent in Charlotte, we see a convincing drop between the CV scores of the best fitting models found during the model search and the best core-plus-four model from the previous section.
In both cities, we see relative humidity appears in many of the top models, although in Charlotte it tends to appear in an interaction.
Hpbl, which as a main effect did not appear in many of the best fitting Charlotte models in the previous section, now appears in all listed models, often in an interaction.
We further notice that many of the interactions in these top models include include a core variable such as wind speed or downward short wave radiative flux, and a planetary boundary layer height or pressure variable.  
Interestingly, few of these interactions include air temperature.  
We also notice that just one of these top models contains an interaction with the precipitation indicator, which may suggest that the presence of precipitation, regardless of other variables, is enough to discourage the most extreme ozone events.

\begin{table}[ht!]
\begin{center}
\caption{The covariates and interactions in the best five models in the automated model search applied to the Atlanta (top) and Charlotte data (bottom). Interactions between continuous covariates are indicated with a `$\times$'. The four core main effects were also included in all models, but do not appear in the covariate list.}
\begin{tabular}{rllll}
  \hline
Rank & CV &  &  &  \\
  \hline
Atl C+4 & 0.5046 & \multicolumn{3}{c}{best core-plus-four model} \\ \hline
Atl 1 & 0.4812 & rel hum & wnd spd$\times$pres chg & wnd spd$\times$ht tropo \\
& & dswrf$\times$NE wnd & dswrf$\times$pres & dswrf$\times$hpbl.max \\
\hline
Atl  2 & 0.4823 & hpbl min & rel hum & ht tropo \\
& & dswrf$\times$NE wnd & wnd spd$\times$hpbl min & N wnd $\times$precip \\
\hline
Atl  3 & 0.4836 & pres & wnd spd$\times$NE wnd & wnd spd$\times$pres \\
& & wnd spd$\times$ht tropo & dswrf$\times$pres & rel hum$\times$NW wnd\\
\hline
Atl  4 & 0.4837 & wnd spd$\times$ht tropo & dswrf$\times$ht tropo & hpbl min$\times$NE wnd \\
& & wnd spd$\times$hpbl min & dswrf$\times$hpbl min & hpbl min$\times$rel hum \\
\hline
Atl  5 & 0.4868 & rel hum & wnd spd$\times$pres chg & wnd spd$\times$ht tropo \\ 
& & new.cloud$\times$pres & temp$\times$hpbl max & dswrf$\times$hpbl max \\ 
\hline
  \hline
Char C+4 & 0.5412 & \multicolumn{3}{c}{best core-plus-four model} \\ \hline 
Char 1 & 0.5085 & hpbl.max & pres chg & rel hum$\times$lwrf \\
& & wnd spd$\times$hpbl min & dswrf$\times$hpbl.max & hpbl.max$\times$rel hum \\
\hline
Char 2 & 0.5172 & hpbl min & rel hum & hpbl.max$\times$NW wnd\\
& & wnd spd$\times$hpbl min & dswrf$\times$hpbl.max & dswrf$\times$hpbl min \\
\hline
Char 3 & 0.5175 & dswrf$\times$NE wnd & rel hum$\times$lwrf & temp$\times$hpbl min \\
& & wnd spd$\times$hpbl min & dswrf$\times$hpbl min & dswrf$\times$rel hum \\
\hline
Char 4 & 0.5177 & dswrf$\times$pres & hpbl.max$\times$pres & hpbl.max$\times$ht tropo \\
& & rel hum$\times$lwrf & wnd spd$\times$hpbl min & dswrf$\times$new.cloud \\
\hline
Char 5 & 0.5181 & dswrf$\times$NE wnd & dswrf$\times$pres & hpbl min$\times$pres \\ 
& & hpbl min$\times$ht tropo & new.cloud$\times$E wnd & wnd spd$\times$hpbl min \\ 
\hline
   \hline
\end{tabular}
\label{tab:SAmodelsAtlChar}
\end{center}
\end{table}

\section{Summary and Discussion}
In this work, we present an atypical multivariate EV study.  
Rather than aiming to assess the probability associated with rare events, we use EV methods to learn about the processes which lead to extreme behavior.  
Specifically, we use the framework of multivariate regular variation to find functions of covariates which exhibit strong tail dependence with the response.  
We employ a multistep data mining procedure where each step built on what was learned from the previous one, and which culminates in an automated model search procedure.
The unique aspect of this study requires novel considerations for extremes such as which tail dependence summary measures are suitable for optimization and the implementation of a smooth threshold.
Fitting and performing cross validation on literally thousands of models required large-scale computational resources.

Our method arose from an applied problem which sought to investigate the meteorological conditions associated with the most extreme ozone levels, which in turn will help atmospheric chemists understand what distinguishes an extreme ozone day from a day with merely high ozone levels.
Not surprisingly, our results show that high air temperature, low wind speed, and high sunlight are influential in producing extreme ozone events.  
However, it appears that covariates besides these likely ones also play a role in distinguishing extreme ozone days.
The models we found that exhibited the strongest tail dependence tended to include interactions between the likely covariates and other covariates, and the fact that the best fitting models include many interactions speaks to the complex relationship between ozone  and meteorological conditions.
Our analysis uncovers local effects such as which wind direction seems to be associated with extreme ozone levels in each of the cities.
We do not end up with a single best model, as there is not likely one unique set of conditions which leads to extreme behavior.
Rather, we provide our collaborating atmospheric chemists with a list of possible contributing factors for further investigation.
That our approach is entirely data-driven and does not include any of the physical and chemical mechanisms which lead to ozone creation means that it provides a view of the drivers of extreme ozone independent from the view given by current atmospheric chemistry models.


There are several avenues for further development.
The meteorological covariates we consider are quite dependent and issues similar to the notion of colinearity warrant further consideration.
We would like to investigate methods which pool information across stations, as the standard errors show that extracting a signal from individual locations is not easy.
A spatial extension to our model could reduce uncertainty and provide additional understanding of how the meteorological drivers of extreme ozone differ over a larger spatial domain. 
As our data mining procedure is closely tied to model selection, one could consider extending the optimization procedure to penalize for complexity (e.g. LASSO \cite{tibshirani1996}). 
However, since our aim is to tease out potential meteorological drivers of extreme ozone for further study, penalizing model complexity is not a primary concern in the optimization phase.



\section*{Acknowledgements}
This research has been supported by EPA STAR Grant RD-83522861-0.  
Cooley also received partial support from DMS-1243102.
We acknowledge high-performance computing support from Yellowstone (ark:/85065/d7wd3xhc) provided by NCAR's Computational and Information Systems Laboratory, sponsored by the National Science Foundation.  
Clemson University is acknowledged for generous allotment of computing time on its Palmetto cluster. 
The authors thank Chihoon Lee for helpful discussion. We also wish to thank an associate editor and an anonymous reviewer for suggestions and comments.


\bibliographystyle{apalike}
\bibliography{biblio/myrefs}

\end{document}